\documentclass[11pt,letter]{article}
\usepackage{epsfig}
\usepackage{amssymb}

\begin{document}

{\huge
\begin{center}
Probing QCD-instanton induced effects in \\
deep inelastic scattering at HERA\footnote{Talk given at the CICHEP 2001 in Cairo, Egypt}
\end{center}
}

{\small
\begin{center}
G.~W.~ Buschhorn\\
Max-Planck-Institut f\"ur Physik\\
(Werner-Heisenberg-Institut)\\
D-80805 M\"unchen
\end{center}
}

{\small After a brief introduction into the phenomenology of instantons,
results from instanton perturbation theory applied to deep-inelastic
electron proton scattering are summarized. Based on event signatures
derived from Monte Carlo studies, strategies for enhancing instanton
induced events relative to Standard Model deep-inelastic events are
developed. Preliminary results for a search for such events performed with
the H1 detector at HERA are given.}

\section{Introduction}

Localized solutions of non-linear field equations known from classical
cases as solitons were discovered also in gauge theories. In 2 spatial
dimensions, these are the Nielsen-Olesen-vortices and in 3 spatial
dimensions, the Polyakov-'t~Hooft-monopoles. In 4 dimensional Euclidean
space (with $x_4 = -ix_0$) such localized solutions of classical Yang-Mills
fields were discovered by Belavin, Polyakov, Schwartz and Tyupkin; since
these solutions are localized in space as well as in time, they were given
the name pseudo-particles or instantons\cite{ref1}.

For non-abelian gauge fields with finite action, the vanishing of the
Lagrange density i.e. $G_{\mu\nu}$ at $|x| \rightarrow \infty$ does not imply
$A_\mu \rightarrow 0$ but only $A_\mu \rightarrow U^{-1} \partial_\mu U$
(``pure gauge''). The corresponding classical vacuum states are then
classified according to a topological number $n$: the winding number
resp. Chern-Simons number. For topological reasons, there exists an
infinite number of non-equivalent vacua.

While classical transitions between these different vacua are forbidden,
quantum-mecha\-ni\-cal tunneling transitions are possible. Since tunneling
corresponds to a classical path in Euclidean space, solutions with minimum
action are favored. The Euclidean action is minimal for
self-/anti-self-dual fields i.e. $G^a_{\mu\nu} = \pm \tilde{G}^a_{\mu\nu}$,
where $\tilde{G}_{\mu\nu}$ is the dual field strength tensor. The tunneling
then is characterized by the topological charge

\begin{equation}
Q = \frac{1}{32\pi^2} \int d^4x G^a_{\mu\nu} \tilde{G}^a_{\mu\nu} .
\end{equation}

\begin{figure}[h]
\begin{center}
\epsfig{file=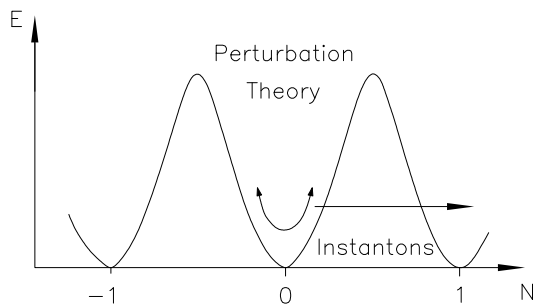,width=12cm}
\end{center}
\vspace{-0.5cm}
\caption{\small Potential energy of classical non-ablian gauge field as function
of the topological charge (Chern-Simons number). \label{fig1}}
\end{figure}

One finds $Q = n_{cs} (t = \infty)-n_{cs} \hspace{0.1cm} (t=-\infty)$,
showing that the tunneling connects different topological vacua. The
tunneling process itself is given the name instanton (see
Fig.~\ref{fig1}). For the tunneling probability, one finds the
characteristic exponential suppression $P \sim \exp(-S_E)$ with the 
Euclidean action $S_E = 8 \pi^2|Q|/g^2$. As was noticed by Ringwald and
Espinosa\cite{ref2}, at high energies the exponential 
suppression of instanton effects in scattering processes can be overcome by
multiple emission of gauge bosons.  Explicitly, the BPST-solution
takes the form\cite{ref3}

\begin{equation}
(G^a_{\mu\nu})^2 = \frac{192 \rho^4}{(x^2 + \rho^2)^2}
\end{equation}

where $\rho$ is a parameter characterizing the size of the instanton.

't~Hooft\cite{ref4} made the important step of coupling light fermions, i.e. massless
quarks, to the gauge field. From the Adler-Bell-Jackiw axial triangle
diagram anomaly, one then has for the axial current $j^5_\mu$ 

\[ \partial_\mu j^5_\mu = \frac{1}{16\pi^2}
G^a_{\mu\nu}\tilde{G}^a_{\mu\nu} . \]

Using (1) it is seen that instantons induce transitions between states of
different axial charge $Q_5(t) = \int d^3 xj^5_0({\bf x},t)$ with

\[ \Delta Q_5 = Q_5(\infty) - Q_5(-\infty) = \int^{+\infty}_{-\infty}dt \int
d^3x\partial_0 j^5_0(x) = \pm 2. \]

This result holds separately for each quark flavor and for $N_f$ flavors
one, therefore, has $\Delta Q_5 = \pm 2N_f$. These topological results,
though derived in Euclidean space, also hold in Minkowski space. Instantons
thus induce non-perturbative hadronic processes violating chirality.

\section{Instantons in deep-inelastic scattering}

Balitsky and Braun\cite{ref5} first showed that the contribution of
instanton $(I)$-induced processes to deep-inelastic electron proton
scattering from a real gluon calculated in Euclidean space and continued to
Minkowski space, rises very rapidly with decreasing Bjorken-$x$. Their
calculations were restricted, however, to $x > 0.3 - 0.35$. A detailed
theoretical\cite{ref6,ref7,ref8} and phenomenological\cite{ref9,ref10}
investigation of deep-inelastic scattering at HERA has been pursued by
Ringwald, Schrempp and collaborators.

\begin{figure}[h]
\begin{center}
\epsfig{file=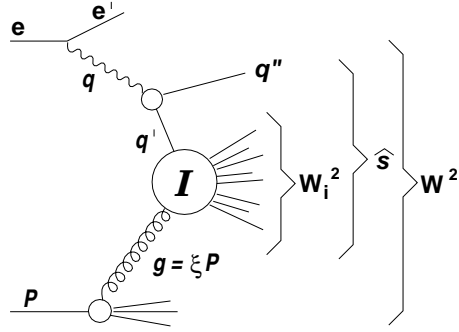,width=6cm}
\end{center}
\caption{\small Leading graph for instanton induced contribution to deep-inelastic
electron proton scattering. It denotes the instanton subprocess with the
variables $Q'^2 = -q'^2 = -(q-q')^2$, $x'=Q'^2/2(g\cdot q')$, $W_i^2 =
(q'+g)^2=Q'^2(1-x')/x'$. \label{fig2}}
\end{figure} 

In perturbation theory, the leading contribution to electron-proton
scattering is photon-gluon fusion (Fig.~\ref{fig2}). The instanton induced
part of the inclusive electron-proton cross section $\sigma^I_{ep}$ is
obtained by an integration of the total quark-gluon cross section
$\sigma^I_{q'g}$ over the photon flux, the gluon density and the quark flux
in the instanton field. $\sigma^I_{q'g}$, which contains all instanton
dynamics, is given by an integral over the ``collective coordinates'' of the
instantons/anti-instantons, i.e. size $\rho$ with the
distribution function $D(\rho)$, separation $R$ and relative color
orientation $U$:

\begin{eqnarray} 
\sigma^I_{q'g} \sim \int d^4Re^{i(p+q')R} \int d\rho D(\rho) d\bar{\rho}
D(\bar{\rho})e^{-(\rho+\bar{\rho})Q'} \cdot \int
dUe^{-\frac{4\pi}{\alpha_s} \Omega(R^2/\rho \bar{\rho}, \bar{\rho}/\rho,
U)}\{...\} 
\end{eqnarray}

with less important contributions collected in \{...\}. $\Omega (..., U)$,
describing the instanton-interaction and incorporating final state gluon
effects, and $D(\rho)$ are known in principle in 
instanton-interaction theory for $\alpha_s(\mu_r)\ln\mu_r \rho \ll 1$ and
$R^2/\rho\bar{\rho} \gg 1$ . $D(\rho)$ has a power law behavior

\[ D(\rho) \sim \rho^{6-(2/3)n_f + 0 (\alpha_s)} \]

which is infrared divergent unless $\rho$ is constrained. In DIS, the
exponential in (3) introduces an effective cut-off $\rho < 1/Q'$ ensuring
a finite integral. 

For large $Q^2$ and the most attractive relative color orientation, 
the integral is dominated by a unique saddle point in $\rho \sim
1/Q'$ and $(R/\rho)^2 \sim 4(x'/1-x')$ which defines a fiducial region in
$x'$ and $Q'$. Taking the limits on $\rho$ and $R/\rho$ from the
UKQCD lattice calculations\cite{ref11} with $\Lambda \frac{n_f}{MS}$ from the
ALPHA collaboration\cite{ref12}, the fiducial region is given by

\begin{displaymath}
\left .
\begin{array}{l}
\rho \stackrel < \sim 0.35 fm\\
\\
\vspace{0.2cm}
\displaystyle{\frac{R}{\rho}} \stackrel > \sim 1.1
\end{array}
\right \}
 \rightarrow 
\left \{
\begin{array}{r}
\displaystyle{\frac{Q'}{\Lambda \frac{n_f}{MS}}} \stackrel > \sim 30.8\\
\\
x' \stackrel > \sim 0.35
\end{array}
\right .
\end{displaymath}

which defines the following ``standard cuts'' to be applied to the experimental
data:

\begin{displaymath}
x \ge 10^{-3}, \hspace{0.3cm} x' \stackrel >\sim 0.35, \hspace{0.3cm} 0.1
\le y \le 0.9, \hspace{0.3cm} Q = Q' \ge 30.8 \Lambda \frac{n_f}{MS}.
\end{displaymath}

In order to suppress contributions from non-planar graphs which are
hard to control, a further cut $Q^2_{min} = Q'^2_{min}$ has been
applied. With these cuts, a HERA cross section $\sigma^I_{ep} \simeq 30 \pm 10$
pb is obtained. Such a cross section is well within the reach of HERA
experiments. 

\begin{figure}[h]
\begin{center}
\epsfig{file=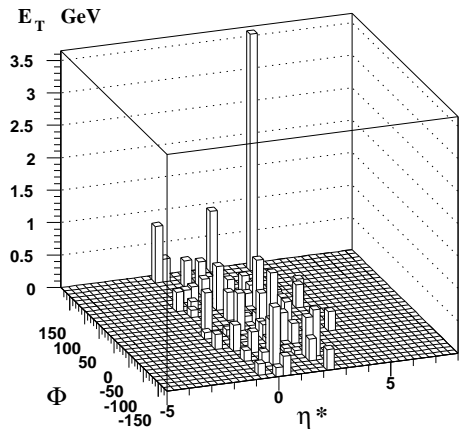,width=6cm}
\end{center}
\caption{\small Distribution of the transverse energy $E_T$ in pseudo rapidity 
$(\eta)$-azimuthal $(\phi)$-plane in the hadronic CMS for a typical
instanton induced HERA-event generated by QCDINS $(x = 0.0012, Q^2 = 66 
\textrm{ GeV}^2, p_T(\textrm{Jet}) = 3.6 \textrm{ GeV})$ after typical
detector cuts. Clearly recognizable are the current jet at $\phi =
160^\textrm{{\small o}}, \eta \simeq 3$ and the instanton band at $0
\stackrel{<}{\sim} \eta \stackrel{<}{\sim} 2$. \label{fig3}}
\end{figure} 

This instanton model has been converted into the Monte Carlo
QCDINS\cite{ref13}. A typical event, shown in Fig.~\ref{fig3}, exhibits the
following characteristic features: 

\begin{itemize}
\item a quark-jet ($q''$ in Fig.~\ref{fig2})
\item a hadron band in the pseudo rapidity-azimuthal angle plane in the
hadronic CMS which is flat in $\phi$
\item high total $E_t$
\item enrichment in heavy flavors ($K, \Lambda, ....$).
\end{itemize}

As pointed out above, the hadronic final state is expected to violate
helicity conservation. 

\section{Monte Carlo Studies}

Monte Carlo generated events have been used to select observables which
discriminate instanton induced events from DIS events. By means of
combining several observables, an enhancement of instanton induced events
is achieved\cite{ref10}.

Standard DIS events have been simulated by two rather different models,
i.e. a matrix element-parton shower model (MEPS) and a color dipole model
(CDM) in order to obtain some information about systematic uncertainties of
Monte Carlos in this extreme region of the phase space. MEPS is based on
RAPGAP which incorporates the $O (\alpha_s)$ QCD matrix element and
models parton emission to all orders in the leading-log approximation by means of parton
showers; the hadronization is performed using the LUND string model as
implemented in JETSET. CDM is using ARIADNE in which gluon emission is
simulated by independently radiating color dipoles and the hadronization
is performed using JETSET.

Instanton induced events have been simulated with the Monte Carlo QCDINS
which consists of the hard instanton process generator embedded in
HERWIG. The QCDINS 2.0 version with its default values $x' < 0.35, Q'^2 >
113$ GeV$^2$ and $n_f=3$ with CTEQ 4L parton densities has been used. The
hadronization is performed using the cluster fragmentation model of HERWIG.

In a first step, the jet with maximal $E_t$ is searched (using a cone
algorithm with $R=0.5$) and identified as the current jet ($q''$ in
Fig.~\ref{fig2}), allowing to estimate the 4-momentum $q''$. $Q'^2$ can be
obtained from $q''$ and from the momentum of the scattered electron. After
removal of the current jet from the event, an instanton band is defined with
$\bar{\eta} \pm 1.1$, where $\bar{\eta} = \sum_h E_{th} \eta_h/\sum_h
E_{th}$. The charged multiplicity $n_b$  in the instanton band is
determined. For simpli\-ci\-ty, the variables $E_{tjet}, Q'^2$ and $n_b$, on
which cuts will be applied, are named ``primary observables''. 

Furthermore, for the particles in the instanton band the transverse energy
$E_{tb}$, the sphericity $Sph$ in the rest-frame of the band and an
observable $\Delta_b$ characterizing the azimuthal anisotropy are defined. We 
define $ \Delta_b = 1 -(E_{out,b},/E_{in, b})$ where $E_{out} =
\textrm{min} \sum_h|\bf{P}_h \cdot \bf{i}| $ and $E_{in} = \textrm{max}
\sum_h |\bf{P}_h \cdot \bf{i}|$ with respect to an axis $i$. For
pencil-like 2 jet-DIS events in photon-gluon-fusion, one expects 
$\Delta_b \rightarrow 1$ and for spherical instanton induced events
$\Delta_b \rightarrow 0$. The variables $E_{tb}, Sph$ and $\Delta_b$ on
which no cuts are applied are named ``secondary variables''.

\section{Experimental results}

The experimental results presented in the following\cite{ref14} have been taken with
the H1 detector at the HERA electron proton collider in 1997 and correspond
to an integrated luminosity of 15.8 $n_b^{-1}$. All results are preliminary.

\vspace{0.5cm}

\begin{figure}[ht]
\begin{center}
\epsfig{file=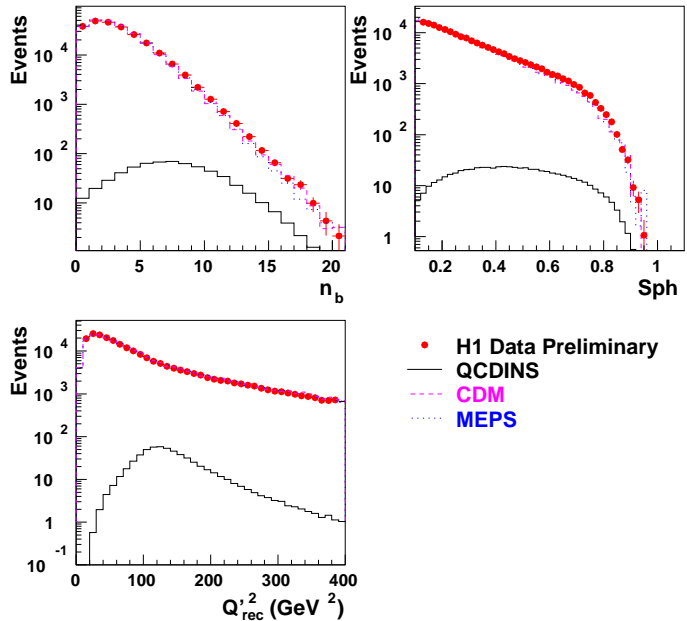,width=9cm}
\end{center}
\caption{\small Observables used to cut (``primary observables''):
comparison of data with Monte Carlo before cuts (see text). \label{fig4}} 
\end{figure} 

In the H1 detector, a central track detector is surrounded by a liquid
argon calorimeter which consists of electromagnetic and hadronic sections
and is covering the forward (with respect to the incoming proton) and
central part of the solid angle. Scattered electrons (positrons actually)
are detected with a backward (with respect to the incoming proton)
electromagnetic calorimeter. Details of the H1 detector can be found
elsewhere\cite{ref15}. 

The event selection (real and Monte Carlo events) required a scattered
electron with $E > 10$ GeV in the backward calorimeter, a vertex within 
$-30 \textrm{ cm} < z < +30 \textrm{ cm}$ of the nominal interaction point
and longitudinal energy conservation within $35 < \sum E-P_Z < 70$ GeV, where $+z$
is the proton beam axis coordinate and the sum is to be taken over all hadrons.

From the scattered electron $Q^2$ and $x$ are reconstructed. The analysis
was restricted to the phase space defined by the electron polar angle
$\Theta_e > 156^{\textrm{o}}, \hspace{0.3cm} 0.1 < y < 0.6$ and $x >
10^{-3}$. The resulting data sample contains 275k events.

\vspace{0.5cm}

\begin{figure}[h]
\begin{center}
\epsfig{file=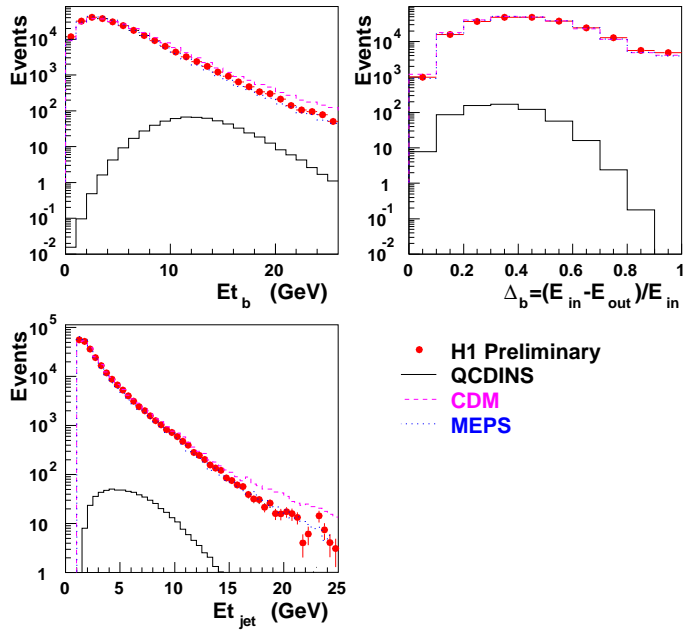,width=9cm}
\end{center}
\caption{\small Observables not used for cuts (``secondary observables''):
comparison of data with Monte Carlo before cuts on primary observables (see
text).\label{fig5}} 
\end{figure} 

Figs.~\ref{fig4} and \ref{fig5} compare H1 data with the Monte Carlos for the different
observables. While the agreement in the primary observables is very good,
some discrepancies are to be noted in the secondary variables $E_{tb}$ and
$E_{tjet}$ at higher values of the variables.

\begin{figure}[h]
\begin{center}
\epsfig{file=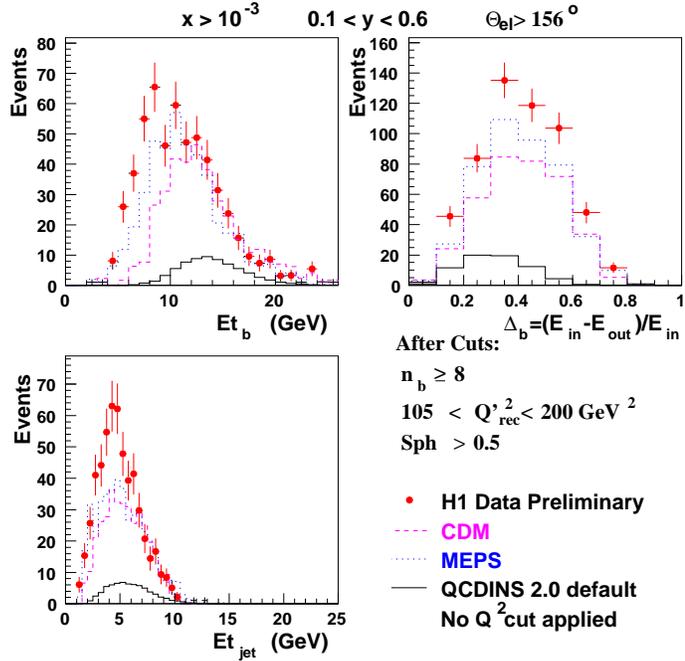,width=9cm}
\end{center}
\caption{\small Observables not used to cut (``secondary observables'') after cuts
on ``primary observables'' \label{fig6}} 
\end{figure}

The effect of cuts on the primary variables on the efficiency $\epsilon_{ins}$ for
retaining instanton induced events, the efficiency $\epsilon_{dis}$ for
retaining DIS background events, and the separation power
$\epsilon_{ins}/\epsilon_{dis}$ for DIS and instanton events has been
systematically investigated. For the combined cuts \mbox{$n_b > 8$},
$105 \hspace{0.1cm} \textrm{GeV}^2 < Q'^2 < 200$ GeV$^2$ and $Sph > 0.5$ one
obtains $\epsilon_{ins} = 11.2\%$ and $\epsilon_{dis} = 0.13 - 0.16\%$
corresponding to a separation power of almost 100. After cuts on the
primary variables 549 events are obtained in the data, while
$363^{+22}_{-26}$ events are expected from CDM and $435^{+36}_{-22}$ 
events from MEPS. The distribution of these remaining events in the 
secondary variables is shown in Fig.~\ref{fig6}. The excess of data over
the DIS expectations has to be contrasted with the difference between the
two different DIS models which is of the same magnitude.

\vspace{0.5cm}

\section{Summary}

Instantons are a fundamental topological feature of non-pertubative
QCD. Instanton perturbation theory supported by lattice QCD calculations
predicts a measurable rate of instanton induced events in deep-inelastic
scattering at HERA.

A strategy for enhancing instanton induced events in DIS events, which is
based on cutting on selected event observables, has been employed
by the H1 collaboration. An excess of data relative to DIS expectation is
observed which is, however, of the same magnitude as systematic differences
of the different DIS model predictions.

The increased data sample available in near future will allow more detailed
investigations.

\vspace{1cm}

\underline{Acknowledgement}:

I thank Shaaban Khalil for the invitation to the conference and
G.~Grindhammer, A.~Ringwald and F.~Schrempp for discussions and a critical
reading of the manuscript.

\end{document}